\documentclass[]{emulateapj}

\def\decdeg {\rlap . {}^\circ}     

\slugcomment{Submitted to Astrophysical Journal}

\shorttitle{HCN and HCO$^+$ in NGC253}
\shortauthors{Knudsen et al.}

\begin{document}


\title{New insights on the dense molecular gas in NGC253 as traced by HCN and HCO$^+$}


\author{K.K. Knudsen\altaffilmark{1}, F. Walter\altaffilmark{1}, A. Weiss\altaffilmark{2}, A. Bolatto\altaffilmark{3}, D.A. Riechers\altaffilmark{1}, \& K. Menten\altaffilmark{2}}


\altaffiltext{1}{Max-Planck-Institut f\"ur Astronomie, K\"onigstuhl 17, D-69117 Heidelberg, Germany}
\altaffiltext{2}{Max-Planck-Institut f\"ur Radioastronomie, Auf dem H\"ugel 69, D-53121 Bonn, Germany}
\altaffiltext{3}{Department of Astronomy and Radio Astronomy Laboratory, University of California, 601 Campbell Hall, Berkeley}


\begin{abstract}
 We have imaged the central $\sim 1$\,kpc of the circumnuclear starburst 
 disk in the galaxy NGC\,253 in the \mbox{HCN(1$\to$0)}, 
 \mbox{HCO$^{+}$(1$\to$0)}, and \mbox{CO(1$\to$0)} transitions 
 at 60\,pc resolution
 using the Owens Valley Radio Observatory Millimeter-Wavelength Array (OVRO). 
 We have also obtained Atacama Pathfinder Experiment (APEX) observations 
 of the HCN(4$\to$3) and the HCO$^+$(4$\to$3) lines of the starburst disk.   
 We find that the emission from the \mbox{HCN(1$\to$0)} and 
 \mbox{HCO$^{+}$(1$\to$0)} transitions, both indicators of 
 dense molecular gas, trace regions which are
 non--distinguishable within the uncertainties of our observations. 
 Even though the continuum flux varies by more than a factor 10 across 
 the starburst disk, the HCN/HCO$^+$ ratio is 
 constant throughout the disk, and we derive an average ratio of
 $1.1\pm0.2$. 
 From an excitation analysis we find that all lines from 
 both molecules are subthermally excited and that they are optically thick.  
 This subthermal excitation implies that the observed HCN/HCO$^+$ line ratio 
 is sensitive to the underlying chemistry. 
 The constant line ratio thus implies that there are no strong abundance 
 gradients across the starburst disk of NGC253. 
 This finding may also explain the variations in $L'_{\rm HCN}/L'_{\rm
 HCO^+}$ between different star forming galaxies both nearby and at high 
 redshifts. 
\end{abstract}



\section{Introduction}

Constraining the physical properties of the dense molecular gas phase 
is important for understanding the relation between the gas properties 
and the star formation process in nearby and high-$z$ galaxies. 
HCO$^+$ has successfully been used as a tracer of dense gas
($n_{\rm H_2} > 10^4$\,cm$^{-3}$) both in star forming regions of the 
Milky Way (e.g., \citealt{christopher05}) and in nearby galaxies (e.g., 
\citealt{nguyen92,brouillet05}).  Recently, it has also been 
detected at high redshifts towards the Cloverleaf quasar \citep{riechers06} 
and towards APM\,08279+5255 \citep{garcia06}. 
HCN is another good tracer of dense gas and has also been observed 
both locally (e.g., \citealt{gao04}) and towards high redshift sources 
(e.g., \citealt{solomon03,vandenbout04,carilli05,wagg05,gao07}).  
\citet{gao04} found that 
the HCN luminosity is correlated with the far-infrared luminosity 
for starforming galaxies over several decades.  
\citet{wu05} found that this correlation holds all the way to the 
smaller scales of the local star forming regions, i.e.\ over 7--8 orders 
of magnitude.  

The number of nearby galaxies in which molecular gas observations allow 
resolving the emission across the dense gas starburst regions is limited 
to only a few.  
At a distance of only about 3.5 Mpc \citep{rekola05}, NGC\,253, a 
member of the Sculptor group of galaxies, is one of the closest 
starbursting galaxies.  This makes it one of the best-suited laboratories
for studies of a circumnuclear starburst and its effect on the
ambient interstellar medium (ISM).
Some early detections of HCN and HCO$^+$ towards the center of NGC253 came 
from \citet{rickard77} and \citet{rickard81}, both using single-dish 
telescopes. 
\citet{paglione95} presented first interferometry maps of HCN 
towards NGC253 using the Nobeyama Millimeter Array. 

We here present Owens Valley Radio Observatory (OVRO) interferometry maps 
and Atacama Pathfinder Experiment (APEX)\footnote{This publication is based on data acquired with the Atacama Pathfinder Experiment (APEX). APEX is a collaboration between the Max-Planck-Institut fur Radioastronomie, the European Southern Observatory, and the Onsala Space Observatory.} 
single-dish mapping 
of both HCN and HCO$^+$, which we use for studying the spatial 
distribution and excitation of the two molecules in the circumnuclear 
starburst of NGC253.  
The paper is structured as follows: 
section~2 presents the observations and data reduction, 
the results are given in \S 3, and in \S 4 we discuss the 
implications both in the context of other nearby galaxies and 
for high redshift sources.  
At a distance of 3.5 Mpc, $1''$ on the sky corresponds to 17.0 pc.


\section{Observations and Reduction}

\begin{figure*}
\plotone{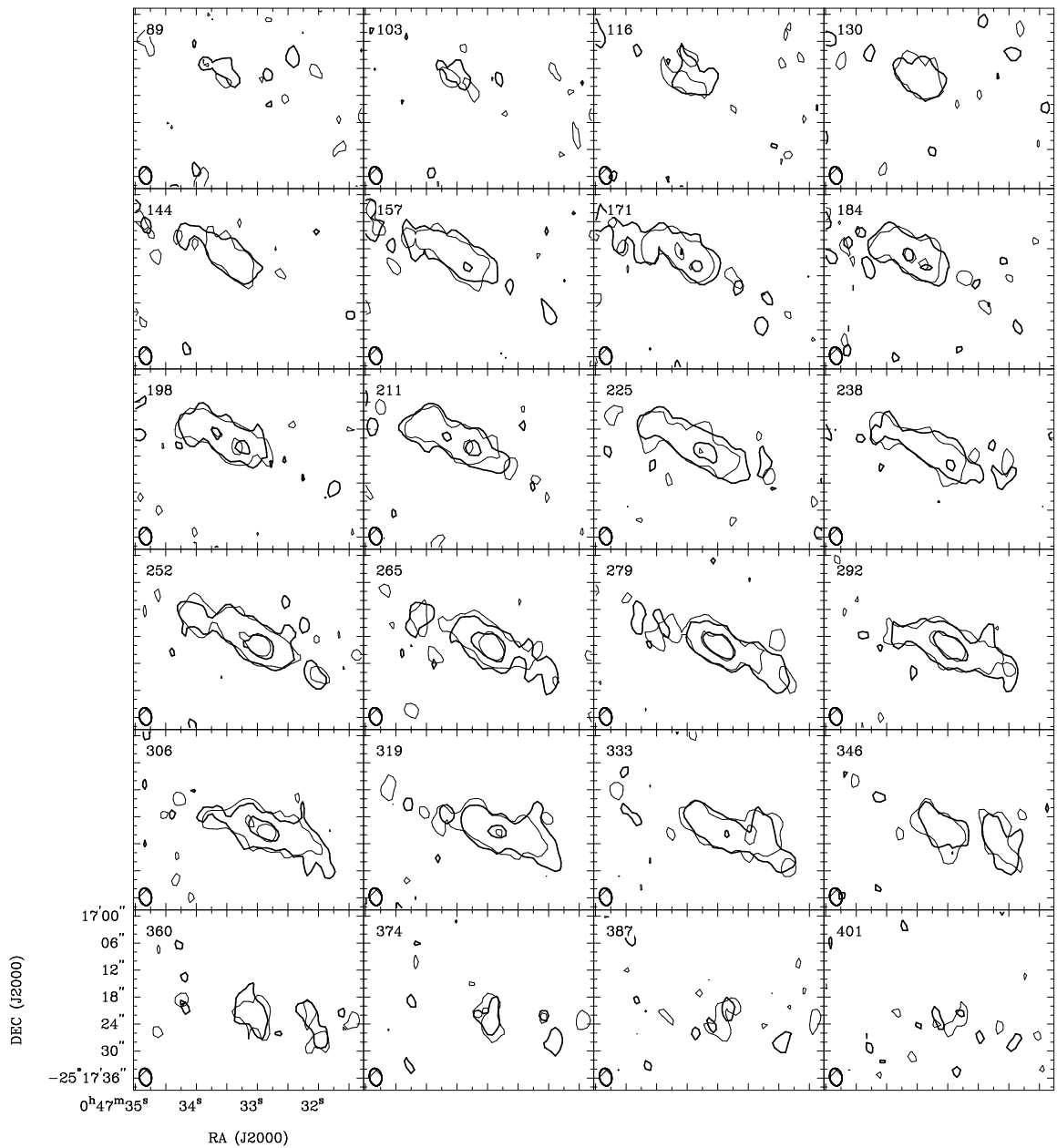}
\caption[]{Channel maps of the HCN (thick contours) and HCO$^+$ (thin 
contours) emission in NGC 253. The beamsize for both maps is identical 
($3.9''\times 2.7''$)  
and two contours are plotted for each cubes at 50 and 200
mJy\,beam$^{-1}$ ($1\sigma$ rms = 22 mJy\,beam$^{-1}$). 
The central velocities for each channel map are given in
the top left corner (in units of km\,s$^{-1}$).
\label{fig:chanmaps}}
\end{figure*}

\subsection{OVRO observations}

We observed NGC\,253 in the HCN(1$\to$0) ($\nu_{\rm rest}$ =
88.63185\,GHz), HCO$^+$(1$\to$0) ($\nu_{\rm rest}$ = 89.18852\,GHz)
and CO(1$\to$0) ($\nu_{\rm rest}$ = 115.2712\,GHz) transitions with
the OVRO Millimeter-Wavelength Array in its C and E configurations.
The HCN and HCO$^+$ observations were obtained simultaneously during four
tracks taken between 1999 November and 2000 February (60 channels,
with a channel width of 4\,MHz, corresponding to
$\sim$13.5\,km\,s$^{-1}$); the CO observations were taken during six 
tracks from 2001 October to 2002 March (120 channels with a width of
2\,MHz $\sim$5.2\,km\,s$^{-1}$); the duration of one track was
typically 5~hr.  The observations for each array were edited and
calibrated separately with the {\sc mma} and the {\sc miriad}
packages.  The {\it uv--}data were inspected and bad data points due
to either poor atmospheric conditions or shadowing were removed, after
which the data were calibrated. Flux calibration was determined by
observing Neptune during most observing runs. For observing runs
without planet observations the flux calibration has been bootstrapped
using the gain calibrator 3C454.3 ($\sim$9~Jy).  These calibrators and
an additional noise source were used to derive the complex bandpass
corrections. The nearby calibrators B0135-247 ($\sim$0.37~Jy),
B0048-097 ($\sim$0.72~Jy), and B0003-066 ($\sim$1.25~Jy) were used
as secondary amplitude and phase calibrators.  We estimate our flux
calibration to be accurate to within 15\%.

For each transition, we produced a data cube using ``robust'' weighting which
was cleaned down to a level of twice the rms noise. This resulted in
beam sizes (and position angles) of 3.79$''\times$2.56$''$ (13$\decdeg$7) and
3.84$''\times$2.64$''$ (8$\decdeg$9) for the HCN and the HCO$^+$ observations,
respectively. As one of the main aims of this study is to compare the
properties of the HCN and the HCO$^+$ emission, we convolved both cubes to a
common resolution of 3.9$''\times$2.7$''$ (10$\decdeg$0), which resulted in an
rms noise of $\sim22$\,mJy\,beam$^{-1}$ in one 4\,MHz channel.  The
resulting beam size for the CO(1$\to$0) observations is 4.9$''\times$2.7$''$
($-1\decdeg$9) with an rms of $\sim65$\,mJy\,beam$^{-1}$ per 2\,MHz channel.
The channels not containing HCN and HCO$^+$ line emission were combined to
create a continuum map at 3\,mm; the resulting rms in this image is
5\,mJy\,beam$^{-1}$.  We note that the resolution of the CO and HCN maps is
comparable to that of the interferometric maps from
\citet{paglione95,paglione04}, while the HCO$^+$ has a higher 
resolution than the Hat Creek Array map from \citet{carlstrom90}.
Furthermore, the resolution of our 3\,mm continuum map is a bit better than
those of \citet{peng96} and \citet{garcia00}. 

As the channel width of the CO(1$\to$0) observations is different from
the HCN(1$\to$0) and HCO$^+$(1$\to$0) observations, we regridded the
CO(1$\to$0) data cube to match the channel width of the other
transitions.  Subsequently, we used the high signal--to--noise ratio 
CO(1$\to$0) data cube to define, in each channel map, the regions that
contain emission (this is called the ``master'' cube).  We used this
master cube to blank both the HCN and HCO$^+$ cubes.  All integrated
maps shown here are based on these blanked cubes and contain the same
areas of emission for all three transitions.

\subsection{APEX observations}

We also obtained observations of the HCN(4$\to$3) ($\nu_{\rm rest}$ = 
354.50547\,GHz) and HCO$^+$(4$\to$3) ($\nu_{\rm rest}$ = 356.73413\,GHz)
transition using the APEX 12m telescope on Chajnantor. For these
frequencies the half--power beamwidth of the antenna is $18''$. 
The observations were carried out during two runs in 2006 September 
in good weather
conditions (typical water columns: 0.7--1mm) using the dual--side--band receiver
APEX-2A. Calibration was achieved using the standard hot--cold--sky
observing scheme roughly every 10\,minutes. For both runs typical system
temperatures were $\sim200$\,K ($T_a^*$). Spectra were recorded using
the Fast Fourier Transform Spectrometer on APEX 
\citep{klein06} providing a bandwidth
of 1\,GHz with an effective resolution of 120\,kHz (0.12\,km\,s$^{-1}$).
The telescope pointing was checked on the nearby carbon star R\,Scl and
found to be accurate to $\sim3''$. The observations were performed in
the standard raster-mapping mode covering the full region where
emission has been detected in the OVRO maps in both transitions. Data 
processing was done using the CLASS software package. In the data processing, 
we dropped all scans with distorted baselines, subtracted linear baselines 
from the remaining spectra, and then rebinned to a velocity resolution of 
20\,km\,s$^{-1}$. 
Finally, temperatures were converted to  $T_{\rm mb}$ using a forward 
efficiency 
and a main beam efficiency of 0.97 and 0.73 respectively \citep{guesten06}.
The conversion from $T_{mb}$ to flux is 32\,Jy\,K$^{-1}$ at 350\,GHz.
We estimate our flux calibration to be accurate to within 15\%.

\begin{figure}
\plotone{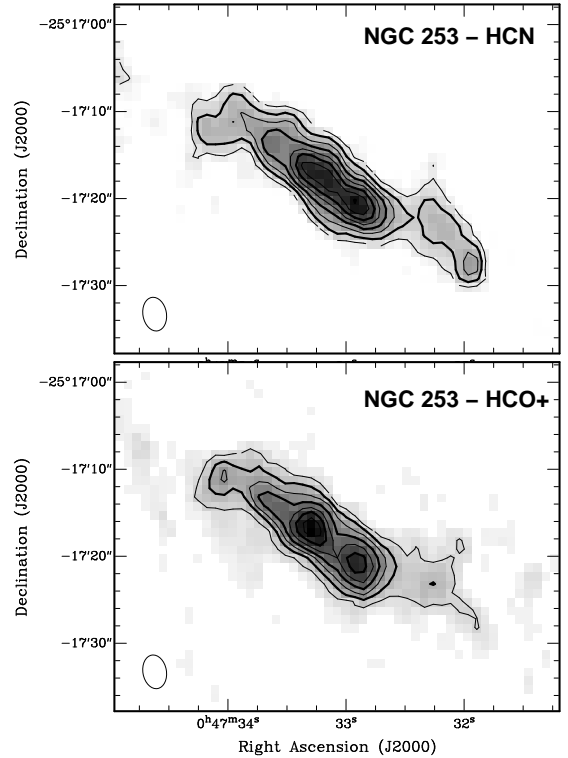}
\caption[]{{\em Top:} Integrated HCN map of NGC\,253. 
{\em Bottom:} Integrated HCO$^+$ map of NGC\,253. 
In both panels, the beamsize ($3.9''\times 2.7''$) is
indicated in the bottom left corner. Contours are plotted starting at
5 Jy\,beam$^{-1}$\,km\,s$^{-1}$ in steps of 5
Jy\,beam$^{-1}$\,km\,s$^{-1}$.  
\label{fig:int_maps}}
\end{figure}


\section{Results}
\label{sect:res}

Figure~\ref{fig:chanmaps} shows the OVRO channel maps for the HCN(1$\to$0) and 
HCO$^+$(1$\to$0) data.  It is striking that their spatial distributions 
are essentially identical.  
Both molecules are tracing dense molecular gas in the circumnuclear 
starburst region of NGC\,253. 

\begin{figure*}
\begin{center}
\includegraphics[scale=0.7,angle=-90]{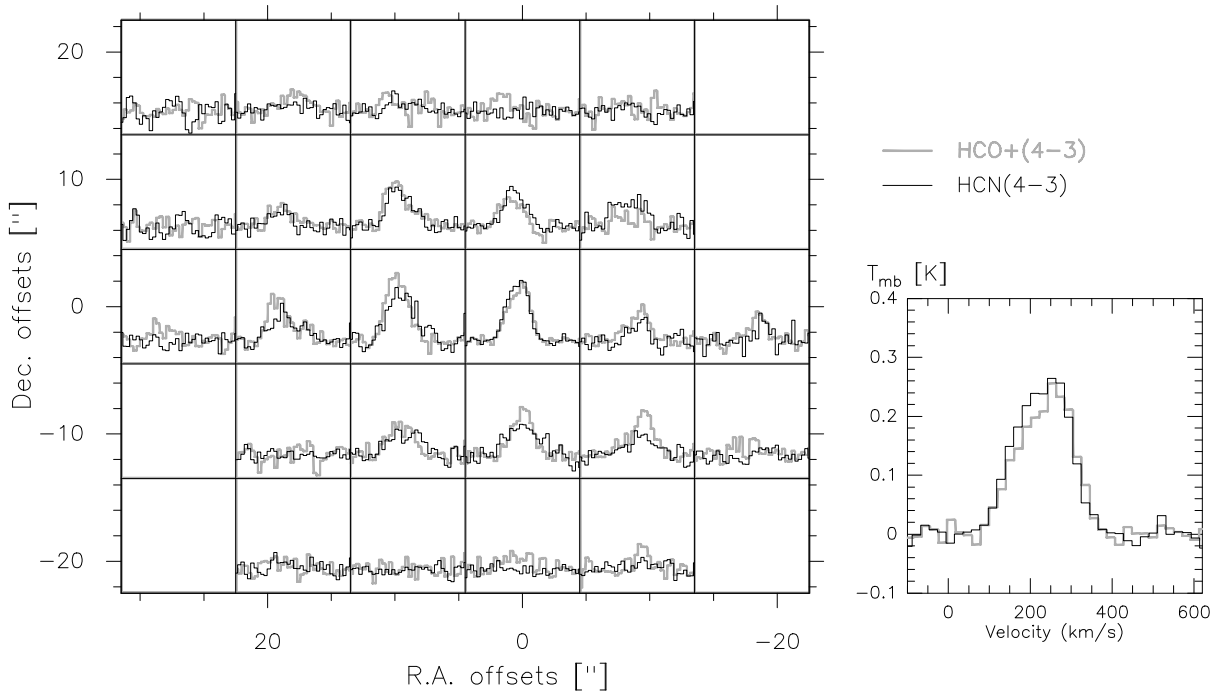}
\end{center}
\caption[]{{\em left:} APEX HCN(4$\to$3) (thin black lines) and HCO$^+$(4$\to$3)
  (thick grey lines) spectra maps towards the nucleus of
  NGC253. Spectra are shown on a velocity scale of -100 to 620
  km\,s$^{-1}$ and on a $T_{\rm mb}$ brightness temperature scale from
-0.1 to 0.4 K. Spectra have been observed on a regular raster with 
half beam spacing ($9''$). {\em right:} Enlargement of the 
HCN(4$\to$3) and HCO$^+$(4$\to$3) spectra towards the central position.  
\label{fig:apexspectra}}
\end{figure*}

Figure~\ref{fig:int_maps} shows the integrated maps of HCN(1$\to$0) and 
HCO$^+$(1$\to$0)  line emission, respectively.  
The slightly different morphology towards the southwest in the 
HCO$^+$(1$\to$0) map compared to HCN(1$\to$0) is likely due to the 
slightly lower signal--to--noise ratio in the map.  
The HCN(4$\to$3) and HCO$^+$(4$\to$3) spectra from APEX are shown in 
Figure~\ref{fig:apexspectra}. 
The line profiles are very similar at all observed positions, and 
the rotation of the circum\-nuclear gas is nicely visible in both transitions. 
The integrated fluxes  for all transitions are given in Table~\ref{tab:results}
along with the line luminosities, which 
we derive using the equations given in \citet{solomon97}. 
The values for HCN and HCO$^+$ ground transitions are consistent with 
previous results from \citet{nguyen89} and \citet{nguyen92} 
for both molecules and \citet{paglione95} for HCN.  

The fluxes and luminosities given in Table~\ref{tab:results} imply a 
HCN(1$\to$0)/CO(1$\to$0) flux ratio of 0.084 or a luminosity ratio of 0.14, 
and a HCO$^+$(1$\to$0)/CO(1$\to$0) flux ratio of 0.077 or a luminosity ratio 
of 0.13, which is similar to other starbursts 
(e.g., \citealt{nguyen94,jackson95,sorai02,gao04,nakanishi05}).
If only measuring the CO flux in those pixels where there is HCN emission, 
the CO flux decreases by 12\%, indicating that there is relatively 
little diffuse CO emission present around the bright nucleus. 

For all three molecules we measure the peak surface brightness at the 
central position: 
CO(1$\to$0), 2.4\,Jy\,beam$^{-1}$,;
HCN(1$\to$0), 0.49\,Jy\,beam$^{-1}$; 
and HCO$^+$(1$\to$0), 0.38\,Jy\,beam$^{-1}$.  
From this we determine the peak brightness temperature 
using the conversions 6.95\,Jy\,K$^{-1}$ at 115.27\,GHz and 14.9\,Jy\,K$^{-1}$ 
at 89\,GHz for our observations 
and derive the following temperatures: 17\,K for CO(1$\to$0), 7.3\,K for 
HCN(1$\to$0), and 5.7\,K for HCO$^+$(1$\to$0).   
Thus, our CO data set a lower limit to the kinetic temperature $>20$\,K 
(Planck temperature). 
The CO(1$\to$0) brightness temperature is in agreement with 
\citet{paglione04}, although it is somewhat lower than the 
36\,K derived by \citet{sakamoto06} at comparable angular resolution. 

\begin{figure}
\plotone{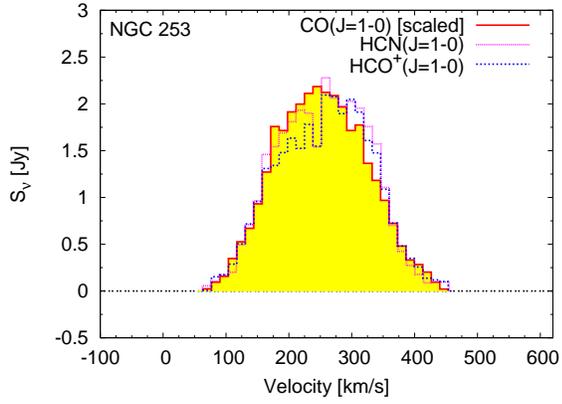}
\caption[]{
Global CO(1$\to$0), HCN(1$\to$0) and HCO$^+$(1$\to$0) spectra of NGC253 
derived from the OVRO data cubes.  
The CO spectrum has been arbitrarily scaled down (by a factor 9.5). 
Each channel is 13.5\,km\,s$^{-1}$ wide.  
The same spectral region is shown as in Fig.~\ref{fig:apexspectra}. 
\label{fig:spectra}}
\end{figure}

For several high redshift sources it has been found that lines from HCN 
appear to be narrower than the CO lines. 
This is the case for the Cloverleaf quasar 
\citep{weiss03,solomon03,riechers06}, APM\,08279+5255 \citep{wagg05,weiss07},
and IRAS\,F10214+4724 \citep{downes95,vandenbout04}.  It has been argued 
that this indicates that the dense gas is more concentrated around the 
nucleus.  Even though NGC253 does not match these high redshift galaxies 
in far-infrared luminosity, we extract the spectra for CO(1$\to$0), 
HCN(1$\to$0), and HCO$^+$(1$\to$0) (shown in Fig.~\ref{fig:spectra}). 
Here we find that the line widths are the same for all three lines. 
This agrees with the interferometric channel maps (Fig.~\ref{fig:chanmaps}), 
in that the emission emerges from the same cloud or system of clouds.

\begin{deluxetable}{lccc}
\tabletypesize{\scriptsize}
\tablecaption{Resulting flux and luminosity \label{tab:results}}
\tablewidth{0pt}
\tablehead{
\colhead{Transition} & aperture & \colhead{$I$} & \colhead{$L'$} \\
\colhead{} & \colhead{(arcsec)} & \colhead{(Jy km\,s$^{-1}$)} & \colhead{(K km\,s$^{-1}$ pc$^2$)} 
}
\startdata
HCN(1$\to$0) & 20 & 186 & $9.4\times10^6$ \\
HCN(1$\to$0) & 27 & 208 & $1.1\times10^7$ \\
HCN(1$\to$0) & Full map & 370 & $1.8\times10^7$ \\
HCN(4$\to$3) & 20 & 1338 & $4.5\times10^6$ \\
HCN(4$\to$3) & 27 & 1634 & $5.5\times10^6$ \\
HCN(4$\to$3) & Full map & 2310 & $7.7\times10^6$ \\

HCO$^+$(1$\to$0) & 20 & 178 & $8.9\times10^6$ \\
HCO$^+$(1$\to$0) & 27 & 203 & $1.0\times10^7$ \\
HCO$^+$(1$\to$0) & Full map & 340 & $1.7\times10^7$ \\
HCO$^+$(4$\to$3) & 20 & 1302 & $4.1\times10^6$ \\
HCO$^+$(4$\to$3) & 27 & 1723 & $5.4\times10^6$ \\
HCO$^+$(4$\to$3) & Full map & 2860 & $8.9\times10^6$ \\

CO(1$\to$0) & Full map & 4400 & $1.3\times10^8$

\enddata
\tablecomments{The lines are detected at $20-30\sigma$.  
This is well within the 15\% uncertainty of the flux calibration. }
\end{deluxetable}

\begin{figure}
\plotone{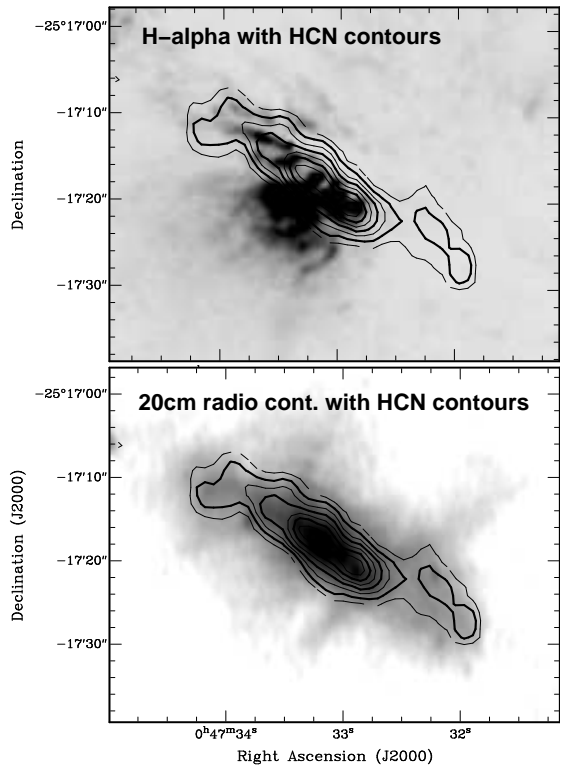}
\caption[]{{\em Top:} H$\alpha$ image (greyscale) with HCN contours 
(contour levels as in Fig.~\ref{fig:int_maps}). {\em Bottom:} Radio continuum
image at 20\,cm, shown as greyscale with a logarithmic scale (again
overplotted with the HCN contours).
\label{fig:halpha_radio}}
\end{figure}

For illustration purposes we also compare the distribution of dense gas 
to H$\alpha$ emission and 20\,cm radio continuum in
Figure~\ref{fig:halpha_radio}. 
Strong H$\alpha$ emission is seen in the dense gas regions traced by 
HCN(1$\to$0) and HCO$^+$(1$\to$0).  In the H$\alpha$ image (from 
\citealt{lehnert95}) the outflow from the central region is very pronounced.  
In the radio 20\,cm map (from \citealt{ulvestad00}), we see that 
the radio emission traces the starburst region as well as the outflow.  
The 20\,cm map compared to high-frequency radio maps is particularly 
useful for comparison with the HCN and HCO$^+$ maps as the resolution 
is similar. 
The region towards the southwest is not seen in the H$\alpha$ image, 
although radio continuum is detected at 20\,cm, indicating that this 
likely is a highly obscured star forming region. 

\begin{figure}
\plotone{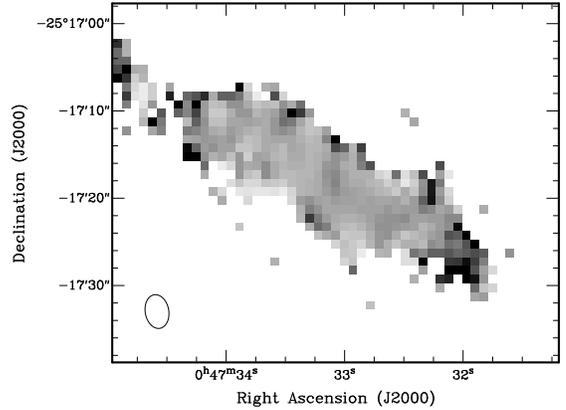}
\caption[]{HCN(1$\to$0)/HCO$^+$(1$\to$0) ratio map of NGC 253. 
Values range from 0.7 (light grey) to 3 (black). 
The beamsize ($3.9''\times2.7''$) is shown in the bottom left corner. 
\label{fig:hcn_by_hco}}
\end{figure}

From the line-free channels of the OVRO observations, we 
derive a 3\,mm continuum image. 
The 20\,cm emission peaks at the center at a flux level of 
0.1\,Jy\,beam$^{-1}$, while the flux level in surrounding regions has 
dropped to about 0.01\,Jy\,beam$^{-1}$. 
In the 3\,mm continuum map, the same is seen, where the flux level also 
peaks around 0.1\,Jy\,beam$^{-1}$ and also falls off by about an order 
of magnitude.  
The continuum emission as seen at 20\,cm is dominated by synchrotron emission
and the 3\,mm emission is likely a combination of free-free emission and dust 
continuum.
A similar behaviour is seen in the Submillimetre Common-User Bolometer Array 
(SCUBA) maps from \citet{alton99}, where 
the submillimeter observations trace the thermal dust emission. 

\begin{figure}
\plotone{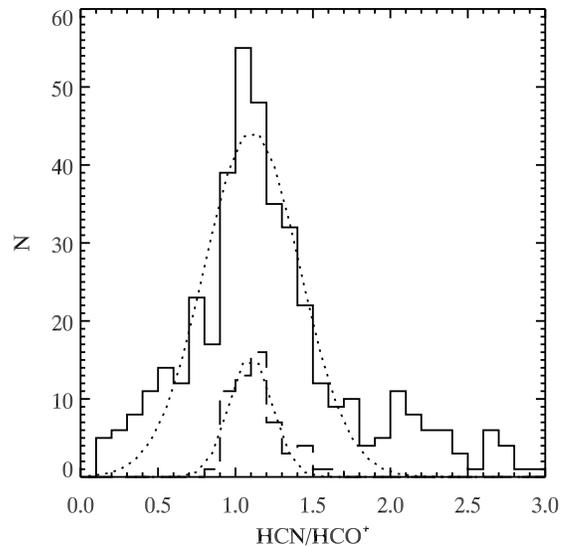}
\caption[]{Histogram showing the distribution in the 
HCN(1$\to0$)/HCO$^+$(1$\to$0) flux ratio map 
(see Fig.~\ref{fig:hcn_by_hco}).  The solid line histogram represents the whole 
starburst region, while the dashed-line histogram shows the distribution 
of the central region, where the continuum emission at 20\,cm, 3\,mm 
and 850\,$\mu$m peaks.  The dotted lines show Gaussian fits to the 
distributions.  
\label{fig:hist}}
\end{figure}

Figure~\ref{fig:hcn_by_hco} shows the HCN/HCO$^+$ ratio map.  Some of the 
pixels along the edges suffer from low signal--to--noise ratio, 
resulting in large 
uncertainties in the ratio in those pixels.  
We construct a histogram of the HCN(1$\to$0)/HCO$^+$(1$\to$0) ratio map, 
shown in Figure~\ref{fig:hist} (solid line), where the tails of the 
distribution are dominated by the low signal pixels.  
We perform a Gaussian fit to the histogram and find an average ratio 
$\langle$HCN(1$\to$0)/HCO$^+$(1$\to$0)$\rangle = 1.1\pm0.3$. 
To check for spatial variations of this value, we also show the 
histogram for the central region only (Fig.~\ref{fig:hist}, dashed line), 
where the continuum level is the highest.  The ratio is $1.1\pm0.2$, while in 
the surrounding regions, where the continuum level has dropped by almost 
a factor 10, it is $1.1\pm0.4$. 
We find no evidence for a change in the HCN/HCO$^+$ ratio across the 
regions with large variation in the continuum level (i.e., radiation field). 


\section{Discussion}

\subsection{HCN and HCO$^+$ excitation}

\begin{figure*}
\epsscale{.95}
\plotone{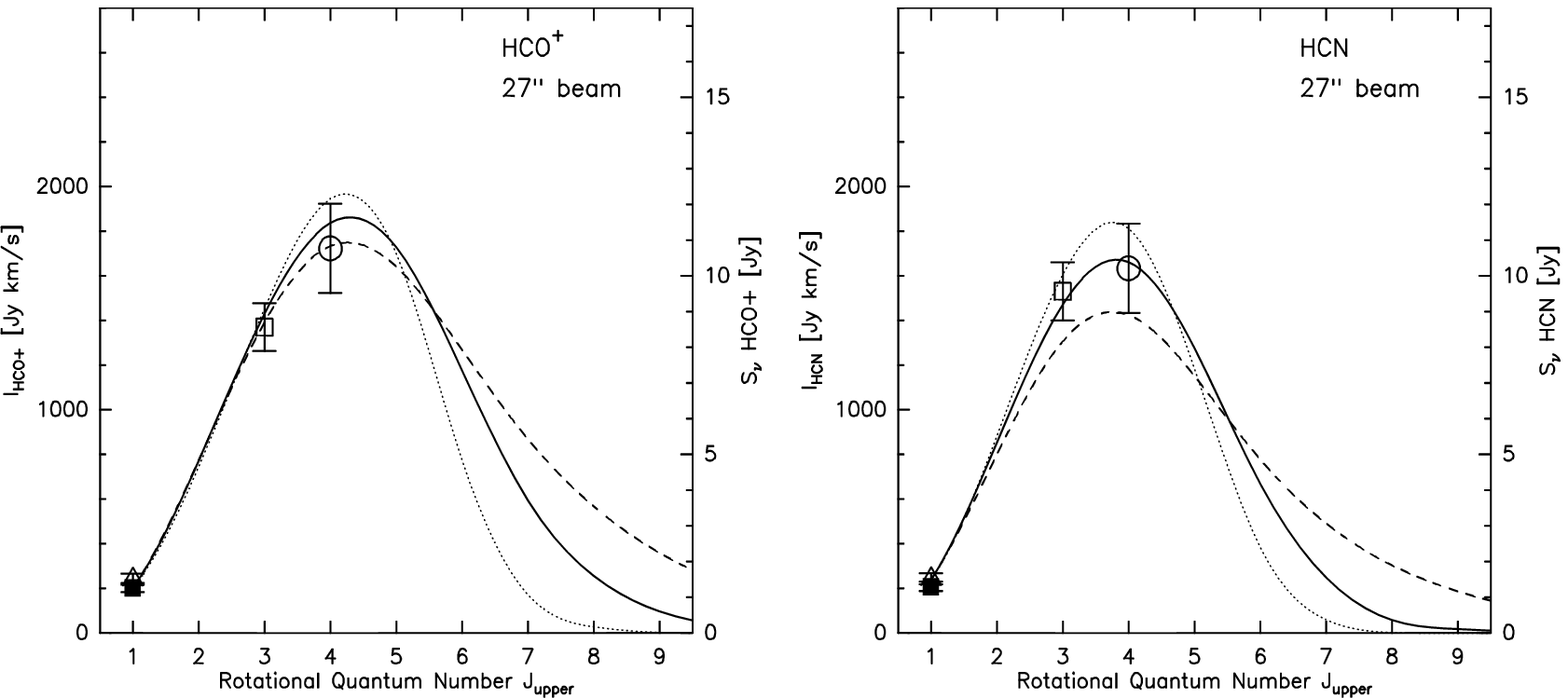}
\caption[]{
HCO$^+$ (right) and HCN (left) line SEDs towards
the nucleus of NGC\,253 in a $27''$ beam. 
The data points are from:  \citet[][triangles]{nguyen92},  
\citealt{paglione97} and \citet{paglione_phd} (HCN and HCO$^+$, respectively, 
open square), 
and this work (Table~1; solid squares and circle).
The lines in both diagrams are selected single-component LVG model
fluxes for [$n({\rm H_2})$, $T_{\rm kin}$] 
combinations; 
solid line: $10^{5.2}$ cm$^{-3}$, 50\,K;
dashed line: $10^{4.9}$ cm$^{-3}$, 150\,K;
dotted line: $10^{5.5}$ cm$^{-3}$, 20\,K. The implied area filling
factors are 1.4\%, 2.0\% and 1.6\% for HCN and 0.9\%, 1.6\% and 1.1\% for 
HCO$^+$.
\label{fig:lvg27sec}}
\end{figure*}

A comparison of the 4$\to$3 to the 1$\to$0 line luminosities of both molecules 
shows that the 4$\to$3
transitions are both subthermally excited independent of the aperture 
chosen (L$'_{\rm HCN(4\to3)}$/L$'_{\rm
HCN(1\to0)}=0.45$, L$'_{\rm HCO^+(4\to3)}$/L$'_{\rm HCO^+(1\to0)}=0.45$).
At a spatial resolution of $20''$, the APEX data is directly
comparable to previous HCN(4$\to$3) measurements from the CSO \citep{jackson95}.
Although the peak brightness temperatures of both
measurements are roughly comparable, our integrated intensity is a
factor of 2 lower than the measurements by Jackson et al (35 vs
71.4 K km\,s$^{-1}$). A similar difference is apparent in the  HCO$^+$(4$\to$3) 
data (31 vs. 50.8 K km\,s$^{-1}$). 
The APEX data in comparison to the Caltech Submillimeter Observatory (CSO) 
data have a much broader bandwidth 
and thus result in a much more reliable baseline subtraction and total fluxes.
Given the quality of our new APEX data we adopt our values for the 
following excitation analysis. 

As the HCN(4$\to$3) and HCO$^+$(4$\to$3) intensities are critical to constrain 
the excitation, we have reinvestigated the gas excitation by applying a series 
of spherical, single-component, large velocity gradient (LVG) models. We
use the collision rates from \citet{schoier05} and relative abundance 
ratios of [HCN/H$_2$] = $5\times10^{-9}$ and [HCO$^+$/H$_2$] =
$1.6\times10^{-9}$ determined for NGC\,253 by \citet{martin06} and
a velocity gradient of 8 km\,s$^{-1}$\,pc$^{-1}$. 
The latter is motivated from CO
studies which indicate that the a large velocity spread is required in 
order to model the high $^{12}$CO/$^{13}$CO line ratios observed towards
NGC253 and other starburst galaxies 
\citep[e.g.][]{aalto91,paglione01,weiss05b}. 
To compare the LVG predicted 
line brightness temperatures to the observations we convert the LVG
line temperatures to integrated flux densities. 
Note that for fixed abundance ratios and a fixed velocity gradient the 
column density per velocity bin scales linearly with the density: 
$N_{\rm mol}/dV \sim X{\rm [mol]}*n({\rm H_2})/(dv/dr)$.

Figure\ref{fig:lvg27sec} shows line spectral energy distributions (SEDs),  
i.e., rotational quantum number versus flux density, 
for HCN and HCO$^+$ for the central 27$''$ beam 
which allows us to also include the HCN(3$\to$2) and HCO$^+$(3$\to$2) from 
\citet{paglione_phd} and \citet{paglione97} to this plot. 
Both SEDs limit the possible range in H$_2$ density to about 
10$^{4.9}$-10$^{5.5}$\,cm$^{-3}$. Models with lower density
fail to reproduce the observed 4$\to$3 flux densities for both molecules.
Similarly, higher densities over-predict the 3$\to$2 and 4$\to$3 lines even
for very cold temperatures of 15\,K. 
The kinetic temperature is 
poorly constrained from the models (see also the LVG models in
\citealt{jackson95}; our Fig.~\ref{fig:lvg27sec}, dotted and dashed lines); 
however, a kinetic 
temperature of 50\,K is the most plausible solution as it is consistent 
with the dust temperature \citep{melo02} and also 
in agreement with models of the CO SED \citep{guesten06}. 
We note that a temperature of $\sim100$\,K, as derived by \citet{bradford03}, 
cannot be ruled out by the model. 
With this, a good match to both line SEDs in our fixed 
chemistry framework is provided by an H$_2$ density of 10$^{5.2}$
cm$^{-3}$, a kinetic gas temperature of $T_{\rm kin}=50$\,K, and an area
filling factor of about 1\% (Fig.~\ref{fig:lvg27sec} solid line). 
All solutions which match the data predict that
HCN and HCO$^+$ lines are {\it optically thick and subthermally
excited}, even in the ground transition of HCN and HCO$^+$. 
For the 50K model we get an excitation temperature $T_{ex}$ of 26K for 
HCN(1$\to$0) and of 42K for HCO$^+$(1$\to$0) and all higher transitions have 
lower $T_{ex}$. 
For a very cold solution of 20K ($n_{H_2} = 10^{5.5}$\,cm$^{-3}$)
the HCO$^+$(1$\to$0) emission would be close to being thermalized.   
Although the area filling factors for HCN and HCO$^+$  differ slightly for 
a given $n_{H_2}$ -- $T_{\rm kin}$ LVG input, the good agreement of the model
predicted SEDs for HCN and HCO$^+$ with the observations suggests that the
emission from HCN and HCO$^+$ arise from a common volume.

\begin{figure}
\plotone{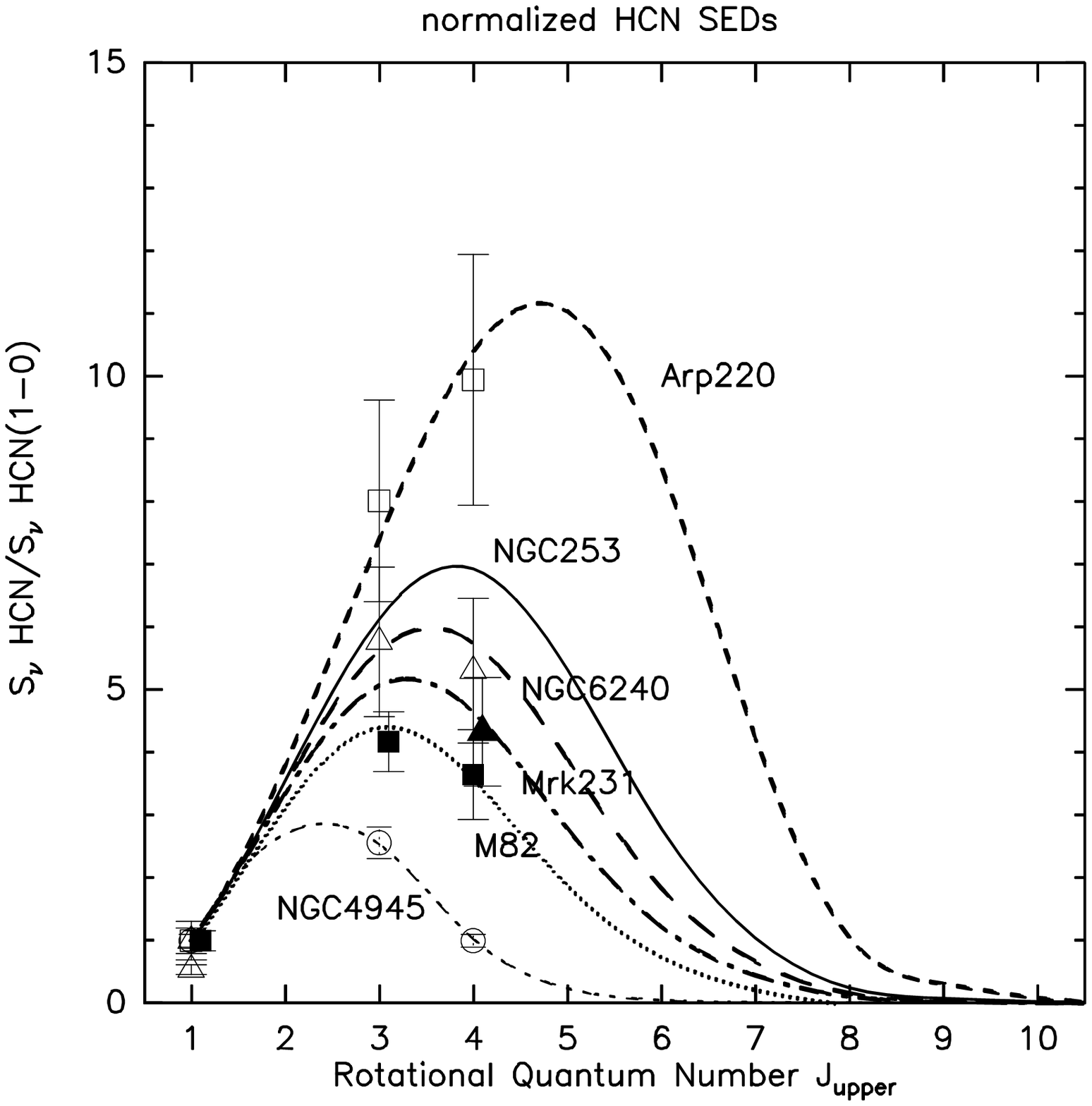}
\caption[]{Comparison of the HCN line SEDs of selected local
  galaxies. The SEDs are shown for NGC253 (this paper; no data points 
  plotted; Fig.\ref{fig:lvg27sec}),  
M82 (solid squares; \citealt{seaquist00}), 
Arp220 (open squares; \citet{greve06} and references therein),
NGC6240 (open triangles; \citet{greve06} and references therein 
[For the SED fit to NGC6240 we have used
the HCN(1$\to$0) flux measurement of \citet{solomon92} as the more recent value
by \citet{greve06} does not provide a good fit in conjunction with the 
HCN(3$\to$2) and HCN(4$\to$3) measurements.]),
Mrk231 (solid triangles; \citet{papado06}),  
and NGC4945 (open circles; \citealt{wang04}). The HCN line SEDs
are normalized by their HCN(1$\to$0) flux density.
\label{fig:normseds}}
\end{figure}

We repeat the analysis for the central position of NGC253 after convolving 
the data to a resolution of 20$''$. 
For our LVG model with $n({\rm H_2})=10^{5.2}$ cm$^{-3}$ and 
$T_{\rm kin}=50$\,K, the 
results are consistent with those derived from 27$''$ resolution. 
The only difference 
is the area filling factor which increases for higher resolution. 
If we apply the same LVG model to the total flux within
the maps covered by the OVRO and APEX observations, the LVG predicted HCN and 
HCO$^+$ (4$\to$3) fluxes are slightly higher than the observed values which 
suggests that the HCN and the HCO$^+$ excitation is somewhat lower in the
outer most regions of the data cubes. At a fixed temperature of  
$T_{\rm kin}=50$\,K, however, a small decrease of the H$_2$ density to 
$n({\rm H_2})=10^{5.1}$ cm$^{-3}$ is sufficient to bring the model
prediction and the observations into agreement. 

In summary, the constant HCN/HCO$^+$ flux ratio across the starburst disk 
would require a highly unlikely combination between varying abundances and 
excitation conditions.  Hence it is safe to conclude that there is no 
significant abundance gradient across the disk.

\subsection{Comparison to excitation conditions in other galaxies}

\begin{figure}
\plotone{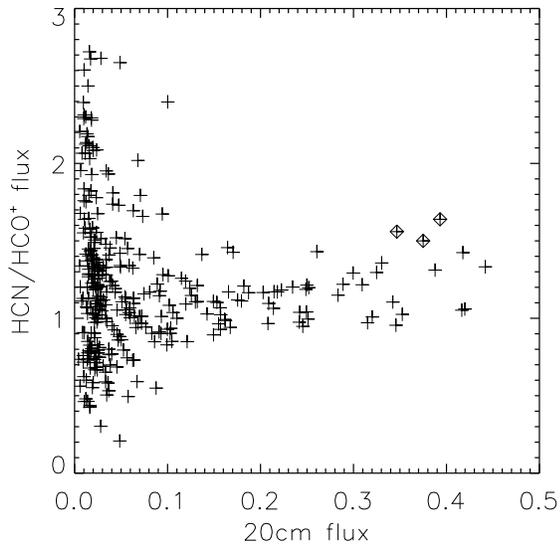}
\caption[]{
Flux ratios of HCN(1$\to$0)/HCO$^+$(1$\to$0) 
plotted as function of 20\,cm flux for each pixel.  The three crosses 
embedded in a diamond are taken from the central position very close 
to the peak of 20\,cm emission, where the angular HCO$^+$ flux distribution 
hints at a divergence from the angular HCN flux distribution. 
\label{fig:ratios}}
\end{figure}

We have used our new HCN observations as well as data from the literature to 
compare the excitation of the dense gas via the HCN SEDs in NGC253, M82,
NGC4945, NGC6240, Mrk231 and Arp220 (references are given in the caption of 
Fig.~\ref{fig:normseds}).  We restrict ourselves here to HCN because 
it provides the most complete data set among tracers of the dense gas for
these active galaxies. All galaxies except Mrk231 have been measured in 
HCN(1$\to$0), HCN(3$\to$2), and HCN(4$\to$3). For Mrk231, only the 
HCN(1$\to$0) and HCN(4$\to$3) have been published in the literature. 
Figure~\ref{fig:normseds} shows a comparison
of the HCN SEDs of the six galaxies normalized by their HCN(1$\to$0)
flux density. This comparison shows that among these sources
NGC\,253 exhibits one of the highest HCN excitations. Given that the
HCN emission in NGC253 is subthermally excited, this comparison 
shows that subthermal HCN is a common property of star forming galaxies. 
The only source
where the HCN excitation exceeds that of NGC253 is Arp220 (see also
\citealt{greve06}; \citealt{papado06}). We note, however, that the HCN line
intensities of Arp220 in \citet{greve06} (which have been used to 
fit the HCN SED of Arp220 in Fig.\,\ref{fig:normseds}) are higher by a
factor of $\sim\,1.5-3$ compared to other studies 
\citep{solomon92,wiedner02,cernicharo06}. 
Therefore, the available data is not conclusive and does
not allow to rule out that HCN is subthermally excited also in Arp\,220.
Interestingly, NGC6240 and Mrk231 show a lower excitation of the dense
gas phase compared to NGC253 (Fig.~\ref{fig:normseds})
despite their much higher FIR luminosities (Fig.~\ref{fig:hcnhco_vs_lir}).

\subsection{Possible interpretation of $L'_{\rm HCN}$/$L'_{\rm HCO^+}$ variations between galaxies} 
\label{subsect:interp}

\begin{figure}
\plotone{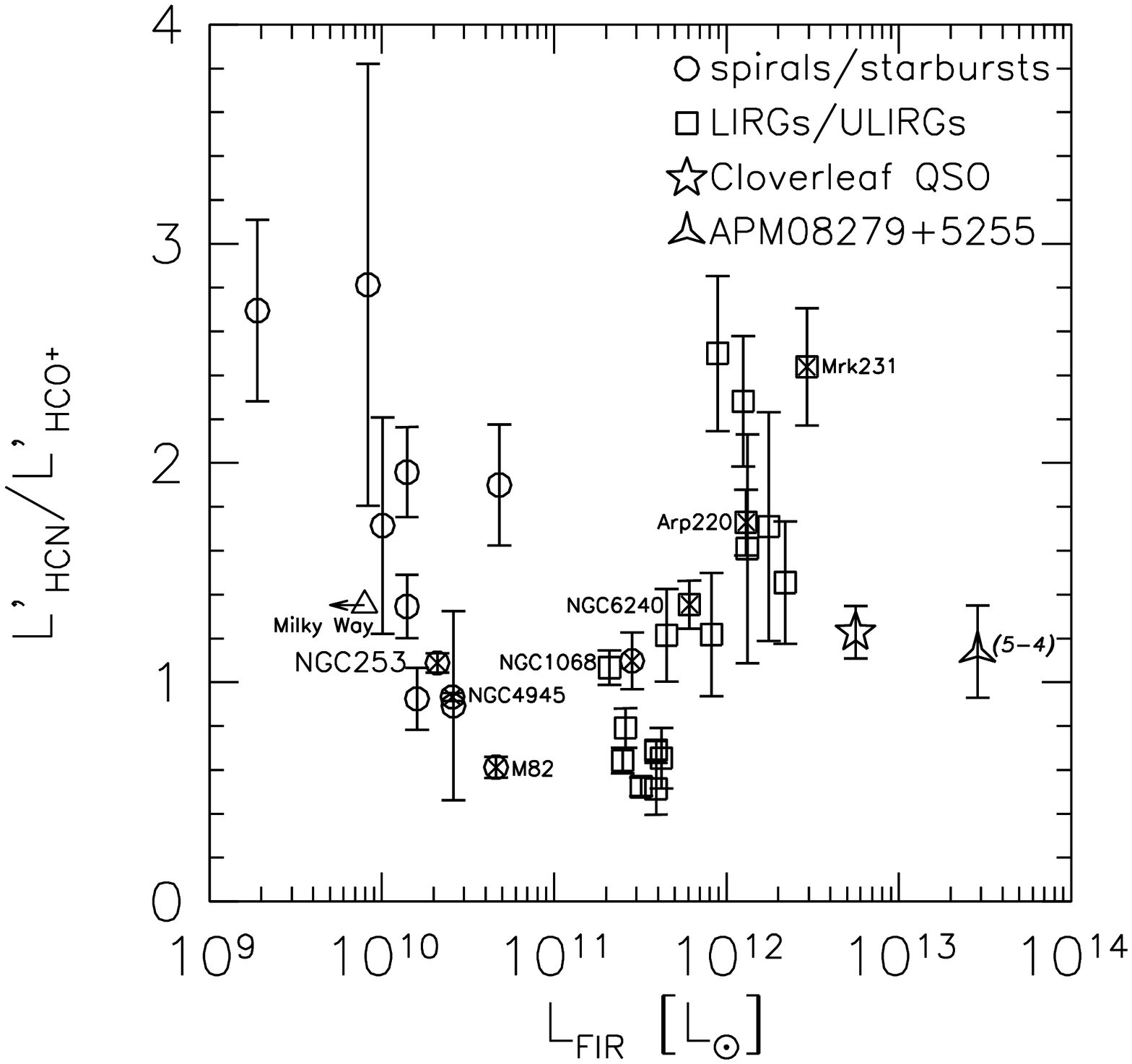}
\caption[]{
HCN/HCO$^+$ luminosity ratio vs.\ the far-infrared luminosity for 
spiral/starbursts 
galaxies \citep{nguyen92,imanishi04,imanishi06,wang04} and LIRGs/ULIRGs 
\citep{gracia06}.  We have 
included our measurements for NGC\,253, the recent high redshift detection 
of HCO$^+$ in the Cloverleaf quasar (\citealt{riechers06}; 
HCN from \citealt{solomon03}) and in the APM08279+5255 (\citealt{garcia06};
HCN from \citealt{wagg05}), 
and indicate the ratio for the circumnuclear 
disk of the Milky Way \citep{christopher05}.  
Note that for all galaxies the $1\to0$ transition was used, expect for 
APM08279+5255 where the $5\to4$ transition was used. 
For the high-$z$ objects, a standard concordance cosmology is used, with 
$H_0 = 73$\,km\,s$^{-1}$\,Mpc$^{-1}$, $\Omega_M = 0.24$, and $\Omega_{\Lambda} = 0.72$ \citep{spergel03,spergel06}. 
\label{fig:hcnhco_vs_lir}}
\end{figure}

While our results show that the HCN/HCO$^+$ ratio is statistically constant 
across the starburst disk, we do inspect the details. 
In Figure~\ref{fig:ratios} we plot the HCN/HCO$^+$ 
flux ratios as function of the 20cm fluxes for each pixel.  
Towards higher 20cm flux, there are three data points which hint at a 
higher HCN/HCO$^+$ ratio
(this increase is due to a relative decrease of the HCO$^+$ emission). 
They correspond to the peak in the 20cm flux.  
This is also towards the peak of the hard X-ray emission as measured with 
{\it Chandra} \citep{strickland00}, where it is possible that there is an 
absorbed active galactic nucleus (AGN) 
present in the very center of NGC253 \citep{weaver02}. 
While these data points are within the scatter, we speculate that the 
small increase in the HCN/HCO$^+$ ratio could be a consequence of the 
presence of an AGN based on the relations suggested by \citet{kohno01}. 
This, however, can only be resolved when future instruments with larger 
resolving power and higher sensitivity will become available.  

The line ratio between HCN and HCO$^+$ has been studied for several classes 
of star forming galaxies, mostly through single-dish observations 
\citep{nguyen92,imanishi04,imanishi06,gracia06}, where ratios have 
been found in the range 0.5-2.7. 
\citet{gracia06} presented a data set of HCN/HCO$^+$ for luminous IR galaxies 
(LIRGs; $L_{\rm IR} > 10^{11}$\,L$_\odot$) and ultra-luminous IR galaxies 
(ULIRGs; $L_{\rm IR} > 10^{12}$\,L$_\odot$).  They present a possible 
correlation between the HCN/HCO$^+$ ratio and the IR luminosity.  If, however, 
other measurements from the literature, including those of less luminous 
star forming galaxies and a recent high-redshift measurement 
\citep{nguyen92,imanishi06,imanishi04,nakanishi05}
are included in such a plot, there remains little evidence for such a 
correlation with $L_{\rm IR}$ \citep[Fig.~\ref{fig:hcnhco_vs_lir}; see also][]{riechers06}. 

\begin{figure}
\plotone{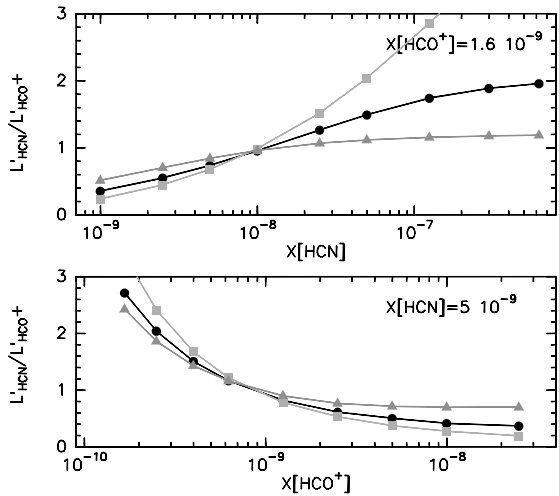}
\caption[]{LVG predicted  $L'_{\rm HCN(1\to0)}$/$L'_{\rm HCO^+(1\to0)}$
  ratios as a function of the HCN (top) and the HCO$^+$ abundance
  (bottom). The lines show LVG models with a H$_2$ density of
  10$^{4.5}$\,cm$^{-3}$ (light grey squares), 10$^{5.0}$\,cm$^{-3}$
  (black circles) and 10$^{5.5}$\,cm$^{-3}$ (grey triangles). The
  kinetic temperature has been fixed to $T_{\rm
  kin}=50$\,K, the adopted velocity gradient is 8 km\,s$^{-1}$\,pc$^{-1}$ for
  all models .
\label{fig:chemistry}}
\end{figure}

The large number of galaxies with $L'_{\rm HCN}$/$L'_{\rm HCO^+}>1$
seems surprising at the first glance since the emission in both lines
is likely optically thick in all galaxies \citep[see also][]{greve06}, 
and the critical densities of HCO$^+$ are lower than those for HCN. 
From this one would expect that the HCO$^+$(1$\to$0) line 
is brighter than the HCN(1$\to$0) line if both molecules trace the same 
gas. However, as both molecules are 
subthermally excited, their line intensity depends on
the underlying chemistry despite their high optical depth. This is 
exemplified in Figure~\ref{fig:chemistry}, where we show the LVG
predicted $L'_{\rm HCN}$/$L'_{\rm HCO^+}$ ratio as a function of reasonable 
HCN and HCO$^+$ abundance for different H$_2$ densities. From the plot
it can be seen that a decrease of the HCO$^+$ abundance can 
indeed explain even the highest observed line ratios independent on
the underlying density of the gas. Alternatively, an increase of the HCN
abundance will also increase the  $L'_{\rm HCN}$/$L'_{\rm HCO^+}$
ratio. The latter process, however, is only effective for H$_2$ gas densities 
below a few times 10$^{5}$\,cm$^{-3}$, as for higher densities  HCN(1$\to$0)
will become thermalized and, therefore, insensitive to abundance variations. 
Given the gas density derived by us for NGC253 (10$^{5.2}$\,cm$^{-3}$),  
both processes can effectively modify the observed line ratio.

A decrease of the HCO$^+$ abundance due to the ionizing field 
produced by cosmic rays has been proposed by \citet{seaquist00}. 
Cosmic rays ionize H$_2$, leading to
the production of H$_3^+$, which reacts with
CO to form HCO$^+$. The abundance of HCO$^+$ is thus affected by ratio of
cosmic--ray ionization rate and gas density.
While a higher ionizing flux favours the production of HCO$^+$, it also
increases the number of free electrons, which
leads to a higher probability for recombination. At a gas density of
$3\times10^4$\,cm$^{-3}$, an ionizing field comparable
to that of the Galaxy may already be strong enough to significantly decrease
the abundance of HCO$^+$ due to dissociative
recombination of H$_3^+$ (e.g.\ Phillips \& Lazio \citeyear{phillips95}). 

Theoretical modelling by
\citet{meijerink05} and \citet{meijerink07} suggest that for modest column densities
$N_H< 10^{22.5}$\,cm$^{-2}$ the presence of an embedded X-ray source, such
as
an AGN, can enhance the HCN/HCO$^+$ and HCN/CO abundance ratios relative
to a pure UV source, such as young massive stars in a starburst. Note,
however,
that for larger column densities the reverse would be true.
There are hints of enhanced HCN/HCO$^+$ ratios in nearby Seyfert galaxies
\citep[e.g.,][]{kohno01}. Nonetheless, similar
enhancements are not found for the far-infrared bright Cloverleaf
QSO \citep{riechers06}.

An alternative explanation for the observed variations of the 
$L'_{\rm HCN}$/$L'_{\rm HCO^+}$ ratio could be that HCO$^+$ and HCN
are not arising form the same volume. From an analysis of the dense gas
properties in Arp220 and NGC6240, \citet{greve06} recently concluded,
that the HCO$^+$ emission is dominantly emitted from regions with 10
times lower density than HCN. In this picture the 
$L'_{\rm HCN}$/$L'_{\rm HCO^+}$ ratio would then be explained by the
relative area filling factors of the denser cores and the less dense
surounding medium in combination with the excitation conditions in
both gas phases. This picture, however, is not supported by our
observation of NGC253, where the excitation of both molecules can be
explained by a single density--temperature combination.

\section{Conclusions}

We have presented maps of HCN(1$\to$0) and HCO$^+$(1$\to$0) 
obtained with OVRO and observations of HCN(4$\to$3) and HCO$^+$(4$\to$3) 
obtained with APEX of the starburst disk of NGC253. 
We find that the spatial distribution of the two molecules is very similar, 
even when compared on a channel-by-channel basis. 
We find the ratio between HCN(1$\to$0) and HCO$^+$(1$\to$0) is $1.1\pm0.2$ 
across the whole circumnuclear starburst region independent of the strength 
of the central radiation field.  
The emission from the \mbox{HCN(1$-$0)} and 
\mbox{HCO$^{+}$(1$-$0)} transitions, both indicators of the 
dense molecular gas, trace regions that are
non--distinguishable within the uncertainties of our observations. 

From an excitation analysis we find that both molecules are subthermally 
excited and that their emission lines are optically thick.  
The constant line ratio implies that there are no strong abundance gradients 
across the starburst disk. 
A consequence of both molecules being subthermally excited is that the 
line intensity is sensitive to the underlying chemistry despite their 
high optical depth.  This finding may also explain the variations 
in $L'_{\rm HCN}/L'_{\rm HCO^+}$ between different star forming galaxies.  

The advent of the Atacama Large Millimeter Array (ALMA), 
Expanded Very Large Array (EVLA), and Square Kilometer Array (SKA) 
will open up improved prospects for 
observations of dense molecular gas phase in galaxies through molecular 
line transitions such as HCN and HCO$^+$ at low and high redshifts. 
As both HCN and HCO$^+$ appear to trace the same dense gas regions 
where the formation of the stars are 
taking place, 
observations of 
these two molecules will be important probes in future 
projects aimed at studying galaxies during phases of the build-up 
of their stellar mass.  




\acknowledgments

The Owens Valley Radio Observatory is supported by the National Science
Foundation through grant AST 99-81546.
We thank the referee, Tim Paglione, for useful comments that helped 
improved the text. 
We thank Michael Dahlem and Rolf G\"usten for useful discussions. 
We thank Matt Lehnert and Jim Ulvestad for making their H$\alpha$ image 
and 20\,cm radio continuum data, respectively, available to us. 
D.A.R.\ acknowledges support from the Deutsche Forschungsgemeinschaft 
Priority Programme 1177.

\end{document}